\documentclass[aps, prb, twocolumn, superscriptaddress, citeautoscript, showpacs]{revtex4-2}
\usepackage{natbib}
\usepackage[dvips]{graphicx}
\usepackage[colorlinks, linkcolor = blue, citecolor = blue]{hyperref}
\usepackage{float}
\usepackage{graphics}
\setcitestyle{super}
\usepackage{bm}
\usepackage{amsmath}
\usepackage{array}
\usepackage{bm}

\begin{document}

\title{Collinear Rashba-Edelstein effect in non-magnetic chiral materials}

\author{Karma Tenzin}
 \affiliation{Zernike Institute for Advanced Materials, University of Groningen, Nijenborgh 4, 9747AG Groningen, Netherlands}
  \affiliation{Department of Physical Science, Sherubtse College, Royal University of Bhutan, 42007 Kanglung, Trashigang, Bhutan}
\author{Arunesh Roy}
 \affiliation{Zernike Institute for Advanced Materials, University of Groningen, Nijenborgh 4, 9747AG Groningen, Netherlands}
\author{Frank T. Cerasoli}
 \affiliation{Department of Chemistry, University of California, Davis, Davis, CA 95616, USA}
\author{Anooja Jayaraj}
 \affiliation{Department of Physics, University of North Texas, Denton, TX 76203, USA}
\author{Marco Buongiorno Nardelli}
 \affiliation{Department of Physics, University of North Texas, Denton, TX 76203, USA}
 \affiliation{The Santa Fe Institute, Santa Fe, NM 87501, USA}
\author{Jagoda S\l awi\'{n}ska}
 \affiliation{Zernike Institute for Advanced Materials, University of Groningen, Nijenborgh 4, 9747AG Groningen, Netherlands}
\date{\today}

\begin{abstract}
Efficient generation and manipulation of spin signals in a given material without invoking external magnetism remain one of the challenges in spintronics. The spin Hall effect (SHE) and Rashba-Edelstein effect (REE) are well-known mechanisms to electrically generate spin accumulation in materials with strong spin-orbit coupling (SOC), but the exact role of the strength and type of SOC, especially in crystals with low symmetry, has yet to be explained. In this study, we investigate REE in two different families of non-magnetic chiral materials, elemental semiconductors (Te and Se) and semimetallic disilicides (TaSi$_2$ and NbSi$_2$), using an approach based on density functional theory (DFT). By analyzing spin textures across the full Brillouin zones and comparing them with REE magnitudes calculated as a function of chemical potential, we link specific features in the electronic structure with the efficiency of the induced spin accumulation. Our findings show that magnitudes of REE can be increased by: (i) the presence of purely radial (Weyl-type) spin texture manifesting as the parallel spin-momentum locking, (ii) high spin polarization of bands along one specific crystallographic direction, (iii) low band velocities. By comparing materials possessing the same crystal structures, but different strengths of SOC, we conclude that larger SOC may indirectly contribute to the enhancement of REE. It yields greater spin-splitting of bands along specific crystallographic directions, which prevents canceling the contributions from the oppositely spin-polarized bands over wider energy regions and helps maintain larger REE magnitudes. We believe that these results will be useful for designing spintronics devices and may aid further computational studies searching for efficient REE in materials with different symmetries and SOC strengths.
\end{abstract}

\maketitle
\section{Introduction}
Many properties of crystals are driven by their structure and symmetries, which not only allows for explaining existing phenomena but also facilitates predicting yet unexplored effects or functionalities in families of materials that are structurally identical. One of the most intriguing symmetries in nature is chirality which describes objects that cannot be mapped onto their mirror images by rotations and translations, the same as human hands.\cite{true_chirality} Chiral molecules and crystals have two energetically equivalent configurations called enantiomers that are mirror reflections of each other. They may respond differently to external fields which gives rise to a variety of intriguing properties and phenomena, for example, unconventional superconductivity, \cite{chiral_superconductor}, Kramers–Weyl fermions,\cite{ nature_material} magnetic structures (skyrmions), \cite{skyrmion1,skyrmion2} or chirality-dependent spin transport that manifests as chirality-induced spin selectivity (CISS) in organic molecules,\cite{ciss1,ciss3,ciss4} or collinear Rashba-Edelstein effect (REE) in chiral materials.\cite{ tellurium_main, te_gate, arunesh_npj} The observation of these effects ignited vigorous debates on fundamental factors that determine them quantitatively, such as the interplay of chirality and spin-orbit coupling (SOC) which exact role remains to be explored.

In our previous studies, we performed first-principles calculations for chiral Te and used the classical Boltzmann transport theory to estimate the spin accumulation generated via REE. We showed that the current-induced spin accumulation is chirality-dependent and parallel to the applied charge current,\cite{arunesh_npj, karma_arxiv} which is in line with the recent experimental studies reporting gate-tunable charge-to-spin conversion in Te nanowires\cite{te_gate}. Similar effects were also investigated in chiral dichalcogenides and metal disilicides, where itinerant electrons become spin-polarized over macroscopic distances in response to the electric field. \cite{crnb3s6, polycrystalline, japanese_prl} Even though some of these works suggest that the observed spin accumulation is the manifestation of CISS, in analogy to the spin signals measured in chiral molecules, the connection between the collinear REE in solid-state materials and CISS in molecular systems is not yet established. Understanding the key ingredients that contribute to these effects is therefore vital.

Here, we report a computational study of two families of chiral compounds, elemental semiconductors Te and Se, and metal disilicides TaSi$_2$ and NbSi$_{2}$ to compare the current-induced spin accumulation in materials with the same crystallographic structure, but different strengths of SOC. By analyzing crystal symmetries and electronic structures of Te and Se, we identify three main factors that may enhance the current-induced spin accumulation in chiral materials; (i) the presence of purely radial spin-orbit field (SOF), (ii) non-degenerate bands spin-polarized entirely along a specific crystallographic direction, (iii) low band velocities manifesting as flat bands. At the same time, we find that stronger SOC broadens energy regions with large REE rather than increases its magnitude. We then generalize these results to the case of metal disilicides. The paper is organized as follows: in Sec. II, we describe density functional theory (DFT) calculations and the details of the REE implementation in the framework of Boltzmann transport theory. In Sec. III, we provide an extended analysis of the electronic structure and spin transport in Te and Se. Section IV describes the analogous properties of the metal disilicides TaSi$_2$ and NbSi$_2$, and Sec. V summarizes the paper.

\section{Computational Details}
Our \textit{ab initio} calculations were performed using the Quantum Espresso (QE) simulation package.\cite{qe1,qe2} We used the fully relativistic pseudo-potentials to treat the ion-electron interaction,\cite{pslibrary} and Perdew, Burke, and Ernzerhof (PBE) version of the generalized gradient approximation (GGA) functional for each material except for Se where we employed PBE for solids (PBEsol).\cite{pbe, pbe_sol} All the structures were constructed using hexagonal unit cells, containing three atoms for the elemental semiconductors and nine for the metal disilicides. We adopted the experimental lattice constants $a$=4.80 (4.78) \AA\ and $c$ = 6.59 (6.57) \AA\ for NbSi$_2$ (TaSi$_2$) reported in the previous studies\cite{japanese_prl, enantiomorph}. For the semiconductors, the structures were fully relaxed and the lattice constants were optimized to $a$=4.52 (4.50) \AA\  and $c$ = 5.81 (5.05) \AA\  for Te (Se). The atomic positions were relaxed using energy and force convergence criteria set to 10$^{-5}$ Ry and 10$^{-4}$ Ry/bohr, respectively. Brillouin zone (BZ) integrations were performed on the $k$-grids of 16$\times$16$\times$12 for TaSi$_2$ and NbSi$_2$ while 22$\times$22$\times$16 and 16$\times$16$\times$14 were used for Te and Se, respectively. As the orbital occupation scheme, we chose the gaussian smearing of 0.002 Ry (NbSi$_2$/TaSi$_2$), 0.001 Ry (Te), and 0.05 Ry (Se). The electronic structures of the semiconductors were corrected by adding the Hubbard parameters calculated via the ACBN0 approach.\cite{acbn0} The values of $U$ were set to 3.81 eV and 3.56 eV for Te and Se, respectively. SOC was included self-consistently in all the calculations, except for the relaxations.

The Fermi surfaces, spin textures and spin accumulation were evaluated using the open-source python package \textsc{paoflow} which projects the wave functions from the \textit{ab initio} self-consistent calculations onto the pseudo-atomic orbitals to construct tight-binding (TB) Hamiltonians.\cite{paoflow1,paoflow2} These Hamiltonians were further interpolated to much denser $k$-grids of 80$\times$80$\times$60 for NbSi$_2$/TaSi$_2$, 200$\times$200$\times$180 for Te, and 200$\times$200$\times$200 for Se in order to sufficiently converge the integrated quantities. While the post-processing calculations of Fermi surfaces, spin textures and spin transport are fully automatized in \textsc{paoflow}, we will briefly describe here our recent implementation of the current-induced spin accumulation used throughout this work.

Let us consider the probability distribution function describing the occupancy of states $f(\boldsymbol{r}, \boldsymbol{p}, t) = f$. The dynamics of this function in the real and momentum space is described via the classical Boltzmann transport theory. In the steady state and using the relaxation time approximation, the Boltzmann equation reads:\cite{citebook1, citebook2, citebook3}
\begin{equation}
	\boldsymbol{v}\cdot\nabla_{\boldsymbol{r}}f_{o} +\boldsymbol{F}\cdot\nabla_{\boldsymbol{k}}f_{o} = -\frac{\delta f(\boldsymbol{k})}{\tau_{\boldsymbol{k}}}
	\label{eqn:eqn1}
\end{equation}
where $\boldsymbol{v}$, $\boldsymbol{F}$, and $\tau_{\boldsymbol{k}}$ are velocity, generalized force, and relaxation time, respectively. We further assume that the system is isotropic and only elastic scattering is taken into account in Eq. \ref{eqn:eqn1}. As the electron distribution is described by the Fermi-Dirac function, the solution of the equation can be written as:
\begin{equation}
	\delta f(\boldsymbol{k}) = \tau_{\boldsymbol{k}} \left(-\frac{\partial f_o}{\partial E}\right)\boldsymbol{v}\cdot\boldsymbol{F}
	\label{eqn:eqn1_soln}
\end{equation}
The net spin accumulation, hence the magnetic moment density induced in the system, is zero at the equilibrium because the spins at opposite momenta cancel each other. If the external electric field $\boldsymbol{E}$ is applied, the net spin accumulation is generated in the system:
\begin{equation}
	\delta \boldsymbol{s} =  \sum_{\boldsymbol{k}}  \langle\boldsymbol{S}\rangle \delta f_{\boldsymbol{k}}
\end{equation}
By substituting the generalized force $\boldsymbol{F}= -q\boldsymbol{E}$, we get:
\begin{equation}
	\delta \boldsymbol{s} = \sum_{\boldsymbol{k}} \langle \boldsymbol{S}\rangle_{\boldsymbol{k}} \tau_{\boldsymbol{k}} (\boldsymbol{v}_{\boldsymbol{k}}\cdot\boldsymbol{E}) \frac{\partial f_{\boldsymbol{k}}}{\partial E_{\boldsymbol{k}}}
\end{equation}
If we assume the constant relaxation time, we can express the above equation in a compact form,
\begin{equation}
	\delta s^j = \chi^{ji} j_i^A
\end{equation}
where $j_i^A$ is an applied charge current and $\chi^{ji}$ is defined:
\begin{equation}
	\chi^{ji} = -\frac{ \sum\limits_{\boldsymbol{k}}\langle \boldsymbol{S}\rangle _{\boldsymbol{k}}^j \boldsymbol{v}_{\boldsymbol{k}}^i  \frac{\partial f_{\boldsymbol{k}}}{\partial E_{\boldsymbol{k}}} } {e \sum\limits_{\boldsymbol{k}} (\boldsymbol{v}_{\boldsymbol{k}}^i)^2  \frac{\partial f_{\boldsymbol{k}}}{\partial E_{\boldsymbol{k}}}}
	\label{eqn:cisp}
\end{equation}
The tensor elements $\chi^{ji}$ are computed using the TB Hamiltonians, whereas the allowed components of the tensor are determined via the symmetry analysis.\cite{arunesh_npj, karma_arxiv, arunesh_prm}

\section{Elemental tellurium and selenium}

\begin{figure*}
	\centering
	\includegraphics[scale = 1.0]{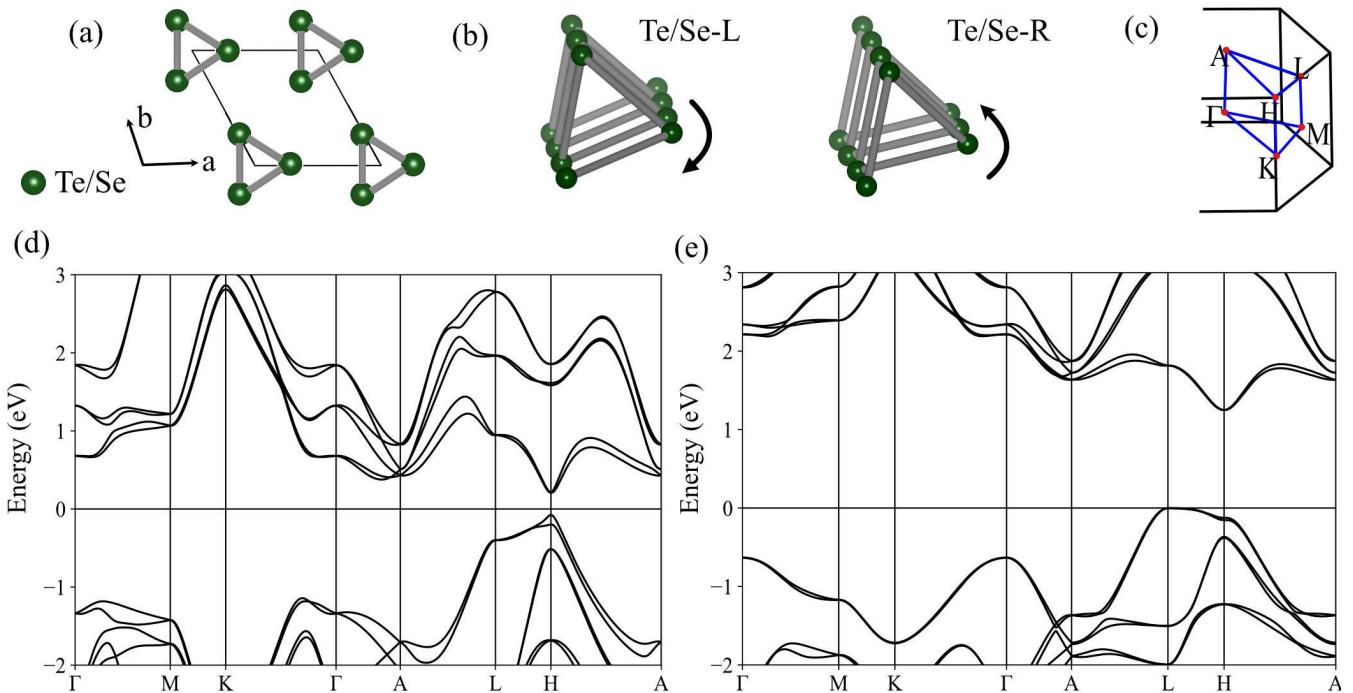}
	\caption{(a) Top view of the crystal structure of trigonal Te/Se. (b) Zoom-in of three Te/Se atoms at one of the corners of the unit cell sketched for the left-handed and right-handed Te/Se. The chirality is determined by the three-fold screw symmetry along the $c$-axis. (c) High-symmetry lines of the Brillouin zone (BZ). (d) The electronic structure of Te calculated along the high-symmetry lines shown in (c). Fermi level is set at the valence band maximum (VBM). (e) The same as (d) for Se. }
	\label{fig:tesebs}
\end{figure*}

\textit{Geometry and electronic structures.}
Elemental Te and Se share the trigonal crystallographic structure described by the space groups P3$_1$21 (152) and P3$_2$21 (154) for the right- and left-handed enantiomers, respectively. The atoms are arranged along the spiral chains running along the $c$ axis that stem from each corner of the \textit{a-b} plane; they interact with each other via weak van der Waals forces (see Fig. \ref{fig:tesebs}(a)). Both enantiomers possess three rotational symmetries C$_2$ whose axes are located within the \textit{a-b} plane, whereas the right-handed and left-handed structures can be distinguished by the threefold screw symmetries $3_1$ and $3_2$ along the \textit{c}-axis (Fig. \ref{fig:tesebs}(b)).

The calculated band structures, illustrated in Fig. \ref{fig:tesebs}(d) and \ref{fig:tesebs}(e), indicate that elemental Te and Se are semiconductors with band gaps of 0.28 eV and 1.25 eV, respectively. The electronic structure of Te is in good agreement with the previous studies,\cite{weyl_node, arunesh_npj, te_gate, tellurium_main, vobornik, sakano} including the value of band gap which is quite close to the measured one (330 meV).\cite{pressure_2} The experimental band gap of Se ($\sim 1.9$ eV),\cite{bandgap_se_3, bandgap_se_2, se_bs} is more difficult to reproduce via standard DFT+U calculations.\cite{weyl_node} However, as both semiconductors are intrinsically $p$-type doped,\cite{pnas1, pnas2, bandgap_se_1} only the valence bands will contribute to the charge and spin transport and our electronic structures are sufficiently accurate to evaluate REE. The bands reveal strong influence of SOC, manifesting mostly as spin-splitting of bands. These features are more apparent in Te, for example, the splitting of the topmost valence band at the $H$-point exceeds 100 meV, one order of magnitude more than in Se ($\sim$ 10 meV). While the exact splitting depends on the $k$-vector, the difference between the materials can be partly explained by the strength of SOI (around 2.7 times larger for the $p$ orbitals of Te, as compared with Se atoms).\cite{fano1971z,martin1971table}

\begin{figure*}
	\centering
	\includegraphics[width=\textwidth]{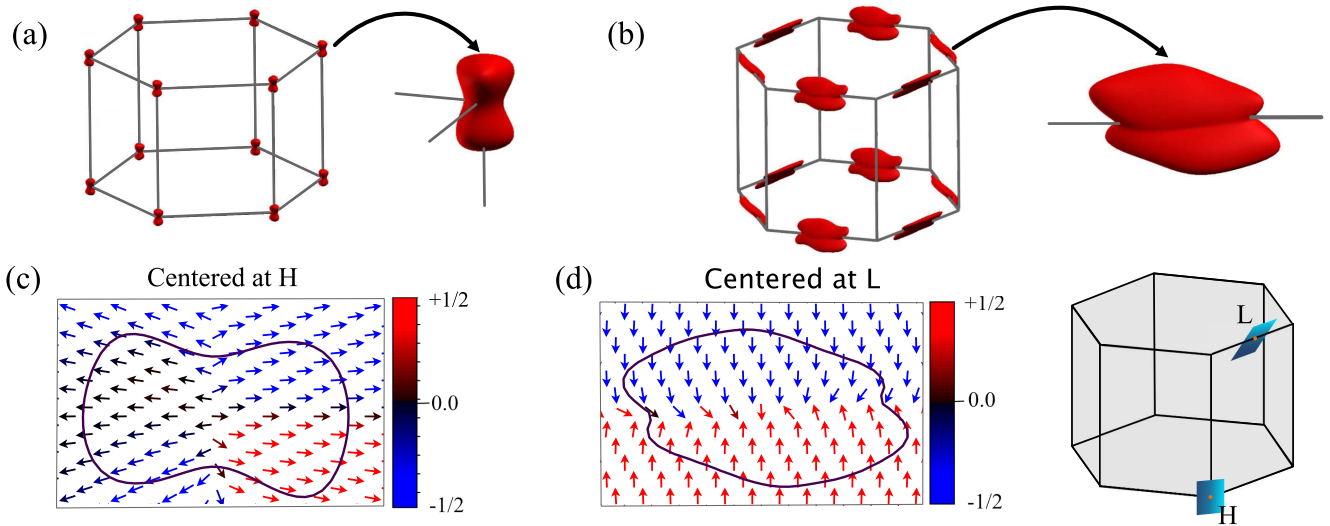}
	\caption{(a) Fermi surface of Te at the energy $E=-30$ meV below the VBM. The zoom-in shows details of the dumbbell-shaped hole pockets that reside at each $H$-point of the BZ. (b) Fermi surface of Se at the energy $E=-10$ meV. All these pockets are centered at the $L$-points. (c) Spin texture of the Fermi surface of Te shown in (a) projected onto the $k_y-k_z$ plane with the $H$-point at the center (see panel (e)). (d) Spin texture of the Fermi surface of Se shown in (b) projected onto the plane of the pocket (see panel (e)). (e) Schematic view of the planes used for the projections shown in (c) and (d).}
	\label{fig:tesespintxt}
\end{figure*}

\textit{Fermi surfaces and spin textures.}
Figure \ref{fig:tesespintxt} shows the Fermi surfaces (FS) and spin textures (ST) calculated a few tens meV below the Fermi level ($E_F$), which corresponds to the slight $p$-type doping observed in experiments.\cite{pnas1, pnas2, bandgap_se_1} In Fig. \ref{fig:tesespintxt}(a), we show the FS of Te consisting of dumbbell-shaped hole pockets at each corner of the BZ.\cite{arunesh_npj} The FS of Se is more complicated, consisting of two intersecting ellipse-like shapes at the L-points, as displayed in Fig. \ref{fig:tesespintxt}(b). Because the electronic and spin transport will be determined mostly by these pockets, we further analyzed the symmetry of SOF around the $H$ ($L$) point of Te (Se). Using the Bilbao Crystallographic Server (BCS),\cite{bcs1,bcs2} we found that the $H$-point of SG P3$_1$21 and P3$_2$21 is described by the D$_3$ wave-vector point group symmetry (WPGS), which can only host the Weyl-type SOC.\cite{zunger} This agrees with the DFT results indicating the radial ST pattern around the $H$-point, as demonstrated by the projection of SOF onto the $k_y-k_z$ plane plotted in Fig. \ref{fig:tesespintxt} (c).\cite{arunesh_npj, karma_arxiv} Similarly, we found that the $L$-point of Se is described by the $D_2$ point group symmetry which allows for the combination of Weyl and Dresselhaus SOF.\cite{zunger} The projection of ST onto the plane of the pocket (Fig. \ref{fig:tesespintxt}(d)) suggests a predominantly radial spin orientation, similar to the case of Te. Moreover, both materials reveal quasi-persistent spin textures which could reduce spin scattering, leading to an increase in spin lifetimes and a robust spin accumulation over large distances.\cite{arunesh_npj, tsymbal_psh, ultrathin_psh}

\begin{figure*}
	\centering
	\includegraphics[scale=1]{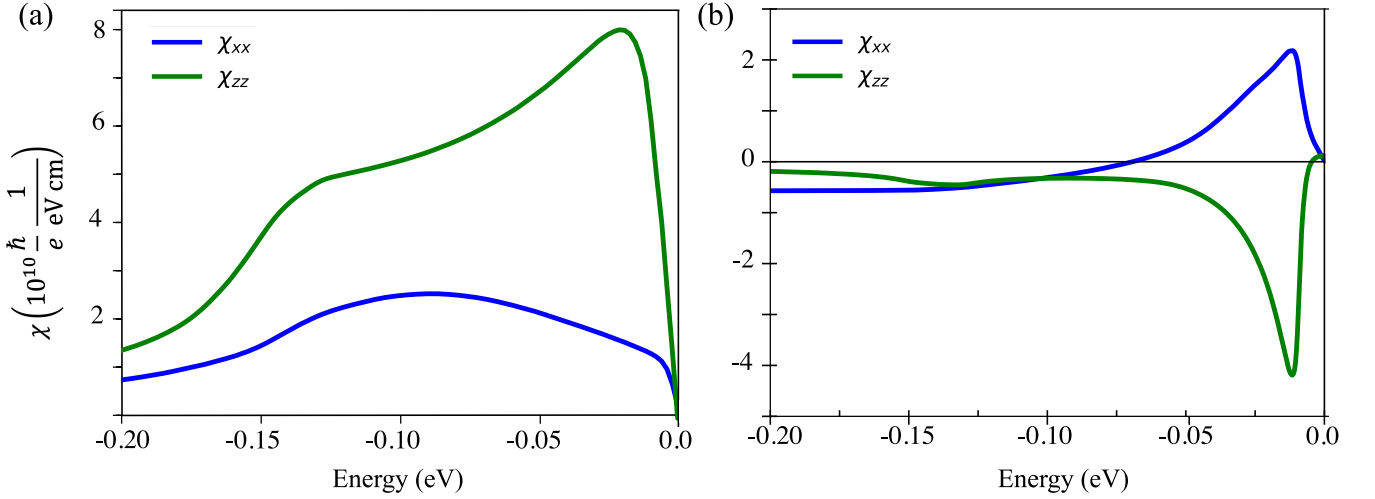}
	\caption{(a) Current-induced spin accumulation vs chemical potential calculated for the left-handed Te. Green and blue lines represent $\chi_{zz}$ and $\chi_{xx}$ components of the REE tensor. The element $\chi_{yy}$ is omitted as it is identical to $\chi_{xx}$.\cite{karma_arxiv} (b) The same as (a), but calculated for Se. $\chi (E)$ of the right-handed enantiomers (not shown) are exactly the opposite due to symmetry.}
	\label{fig:tesecisp}
\end{figure*}

\textit{Spin accumulation.}
We again performed the symmetry analysis using BCS and found that the REE tensor possesses only diagonal elements $\chi_{xx} = \chi_{yy} \neq \chi_{zz}$.\cite{karma_arxiv} Their magnitudes plotted as a function of the chemical potential are displayed in Fig. \ref{fig:tesecisp}(a) and \ref{fig:tesecisp}(b) for the left-handed Te and Se, respectively. Since the electronic structures of left- and right-handed enantiomers are identical and differ only by the sign of ST, the REE magnitudes are exactly opposite, which means that the spin accumulation induced by a given charge current is chirality-dependent. The REE estimated for Te is the largest along the \textit{z} axis with the maximum of 8.00 $\times 10^{10}~\frac{\hbar}{e}\frac{1}{\mathrm{eV cm}}$ at the energy $E\approx-20$ meV below the Fermi level. The values induced by charge currents flowing along the $x$ and $y$ axes are significantly lower, albeit within the same order of magnitude, with the maximum of 2.52 $\times10^{10}~\frac{\hbar}{e}\frac{1}{\mathrm{eV cm}}$ at the energy $E \approx -90$ meV below the VBM. The weaker SOC in Se implies overall lower magnitudes of REE, but the behavior of $\chi (E)$ differs from that of Te, indicating that the exact electronic structure plays a relevant role. For example, the maximal value of $\chi_{zz} = -4.19 \times 10^{10}~\frac{\hbar}{e}\frac{1}{\mathrm{eV cm}}$ at $E\approx-10$ meV is negative. Also, the $\chi_{xx}$ changes the sign close to $E_F$, which does not take place in Te. This means that the sign of spin accumulation could be tuned by switching the direction of the applied electric field.

\begin{figure*}
	\centering
	\includegraphics[width=\textwidth]{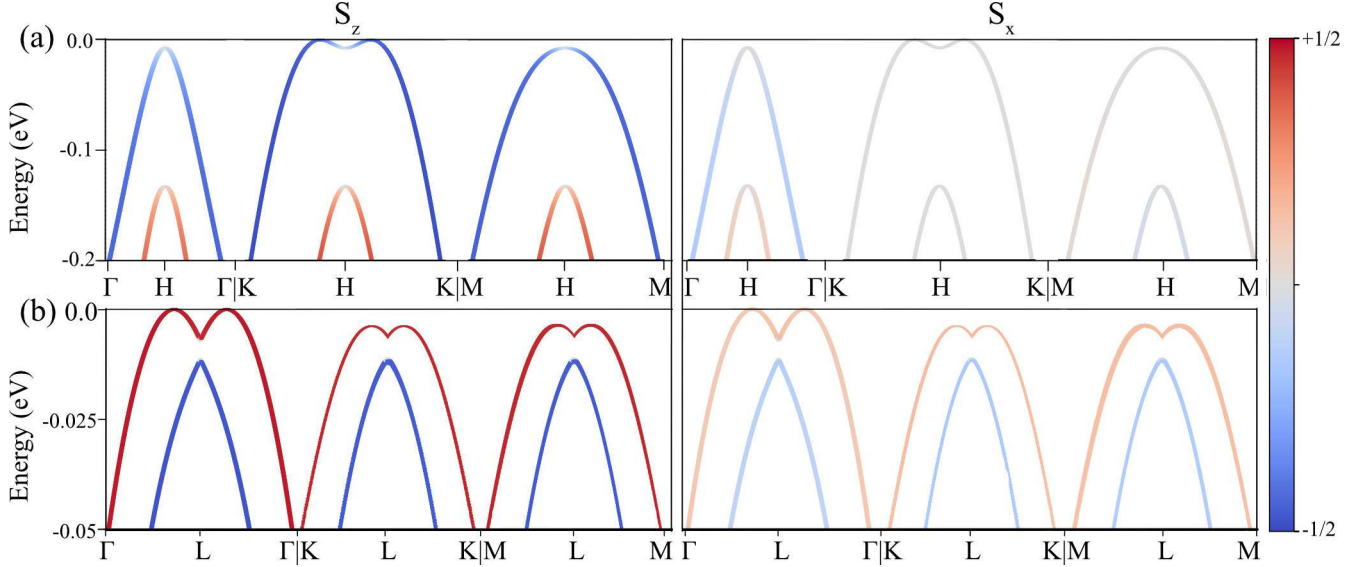}
	\caption{(a) Spin-resolved band structure of the left-handed Te calculated along the relevant high-symmetry lines. The bands are spin-polarized mostly along the $z$-axis. The strengths of the $S_z$ and $S_x$ components are visualized using color (see the legend). (b) The same calculated for the left-handed Se. The spin texture is not polarized either along the $x$- or the $z$-axis, which can be also deduced from Fig. \ref{fig:tesespintxt}(b). Note that the spin-splitting of Se bands is lower than in Te. The spin textures of the corresponding right-handed enantiomers are exactly the opposite due to symmetry (not shown).}
	\label{fig:spin_resolved}
\end{figure*}

\begin{figure*}
	\centering
	\includegraphics[scale=1]{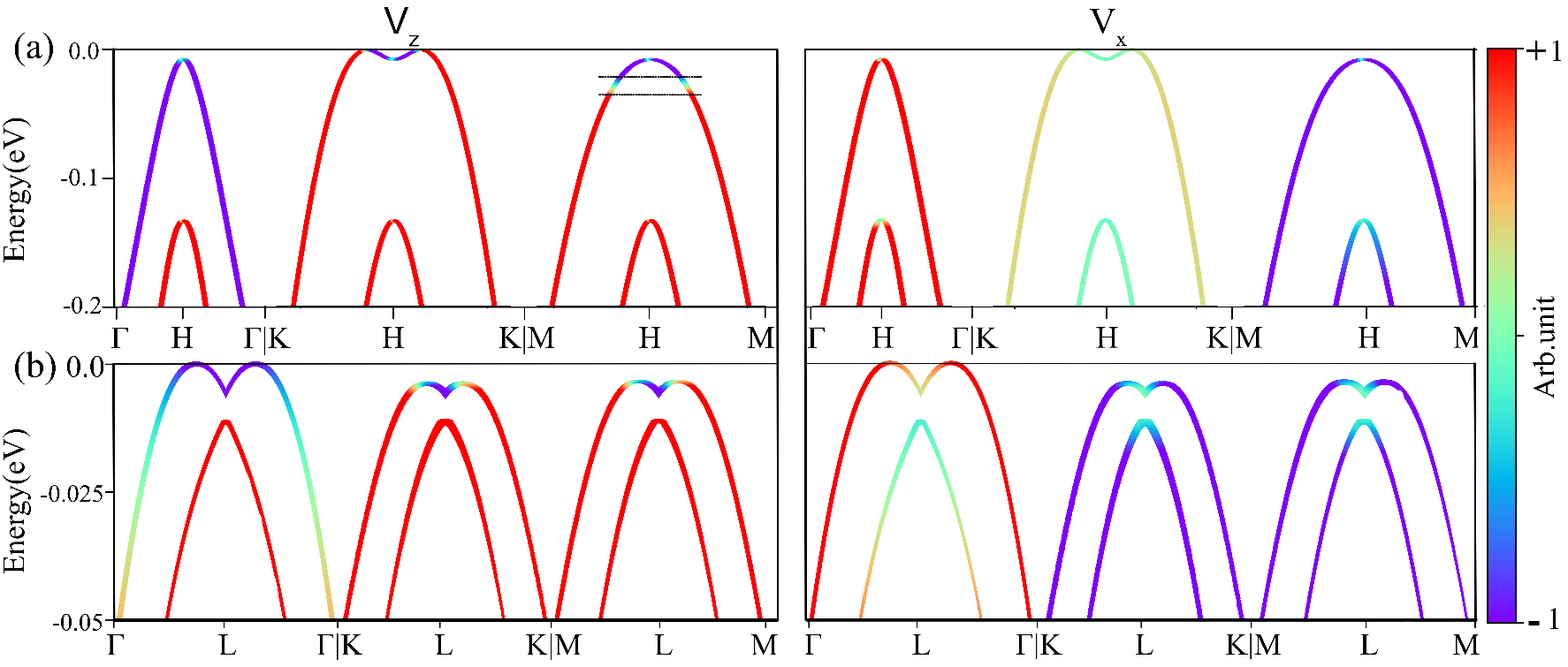}
	\caption{(a) Velocity-resolved band structure of Te-L calculated along the selected lines. The intensities of $v_z$ and $v_x$ components visualized as the color maps are shown in the left and right panels, respectively. (b) The same as (a), calculated for Se-L.}.
	\label{fig:velocity_resolved}
\end{figure*}

To gain more insight into the peculiar $\chi(E)$ plots, we analyze both spin polarization and velocity of bands along possible paths connecting the high-symmetry points in the irreducible BZ (see Fig. \ref{fig:tesebs}(c)). For the energy range considered in Fig. \ref{fig:tesecisp}, only the bands around the $H$-point (Te) and $L$-point (Se) contribute to  spin transport. We will first focus on Te, in which the bands are present along three high-symmetry paths $\Gamma-H$, $K-H$, and $M-H$. Their spin-polarization is illustrated in Fig. \ref{fig:spin_resolved}(a) and we can notice that the spins are oriented predominantly along the $z$ direction. This explains the enhanced spin accumulation along this axis, as the tensor component $\chi_{zz}$ is proportional to the sum of spin expectation values for the $z$ component. The emergence of another band with the opposite spin polarization at $\sim$ -0.13 eV causes a partial cancellation of the generated spin density, which manifests as a decrease in $\chi_{zz}$ starting around this energy level and extending to the energies below. The $\chi_{xx}$ component of the tensor is determined mostly by the $\Gamma-H$ and partly by the $M-H$ path; note that the $S_x$ component of the spin texture is negligible along the $K-H$ line. The intensity of $S_x$ gradually increases from zero to the onset of the next band, resulting in a nearly constant $\chi_{xx}$ with a mild maximum at -0.09 eV followed by a drop below -0.13 eV.

The analysis of the band velocities plotted in Fig. \ref{fig:velocity_resolved}(a) further clarifies the behavior of $\chi(E)$. The $v_z$ component of Te is large and rather uniform in the entire considered energy window and along all the paths. The most interesting features can be found along the $M-H$ path within the energy range between -20 and -35 meV below the VBM. In that region, the velocity $v_z$ is very low which seems to be the cause of the sharp maximum of $\chi_{zz}$ (see Eq. 6). In contrast, the component $v_x$ does not seem to have a large impact on the $\chi_{xx}$ component which hardly varies with $E$. Here, the main contribution comes from the top of the band at the $H$-point, where $v_x$ is very low and causes a fast increase of the $\chi_{xx}$ component below $E_{F}$, despite rather weak $S_x$ in that region (see Fig. \ref{fig:spin_resolved}(a)).

The behavior of the collinear REE in Se can be explained using a similar approach. The major difference, in comparison with Te, is a very small spin-splitting of the valence bands resulting from the weaker SOI; at the $L$-point the bands are separated by only a few meV (see Fig. \ref{fig:spin_resolved}(b)). Moreover, the spin polarization of bands is reversed with respect to Te, which explains the opposite sign of the $\chi_{zz}$ component. The band velocities $v_z$ are similar in Se and Te (Fig. \ref{fig:velocity_resolved}(a-b)), but the former lacks low magnitudes around the VBM and the highest value of $\chi_{zz}$ is twice lower than in Te. Even though low values of $v_z$ appear at the higher binding energies, especially along the $\Gamma-L$ line, their effect is suppressed by the presence of the next valence band with the opposite polarization which causes a fast decrease in $\chi_{zz}$. On the other hand, the behavior of $\chi_{xx}$ seems to be determined by the band velocity ($v_x$). While the sign of the spin accumulation around the $E_F$ is positive and cannot be justified by the spin polarization of bands, $v_z$ and $v_x$ have exactly opposite signs along all the paths shown in Fig. \ref{fig:velocity_resolved}(b), which explains different signs of $\chi_{zz}$ and $\chi_{xx}$. The magnitudes of $\chi_{xx}$ are in general lower because the component $S_x$ of the spin texture is weaker than $S_z$ (Fig. \ref{fig:spin_resolved}(b)).

Overall, the analysis of Te and Se allows us to draw a few important conclusions. While the purely radial spin texture leads to larger magnitudes of REE, as compared to materials with FS endowed with mixed ST patterns,\cite{karma_arxiv} the exact features of $\chi(E)$ curves are sensitive to both the direction of spin polarization and band velocities. The existence of a band where the average spin polarization is along a specific direction leads to the enhancement of the spin accumulation along that direction. Moreover, the bands with low velocities, or ideally flat regions, provide the largest contributions to $\chi (E)$. That means that even materials with relatively low SOC could manifest significant REE, as long as the bands are (nearly) flat. We note, however, that stronger SOC yields greater splittings between oppositely spin-polarized bands, as in the case of Te, so the regions with larger $\chi$ cover wider energy ranges. In the next section, we will show that the spin accumulation in chiral semimetals with more sophisticated Fermi surfaces and spin textures follows essentially the same trends, strengthening these conclusions.

\section{Chiral Disilicides}

\begin{figure}
\centering
\includegraphics[scale = 1.0]{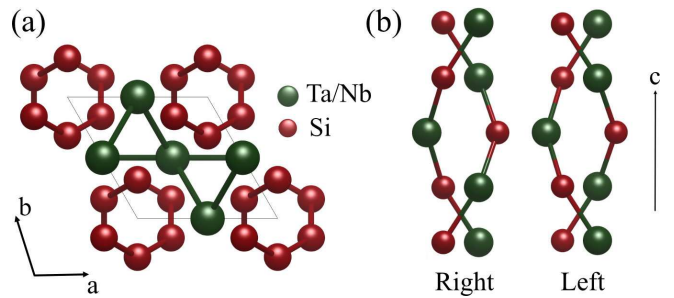}
\caption{(a) Top view of TaSi$_2$ and NbSi$_2$ crystal structures. (b) Side view of their right- and left-handed enantiomers. The chirality is determined by the opposite helicity of the pair of spiral chains running along the $c$-axis (SGs 180 and 181).}
\label{fig:tanbstruct_0}
\end{figure}

\begin{figure*}
	\centering
	\includegraphics[scale = 1]{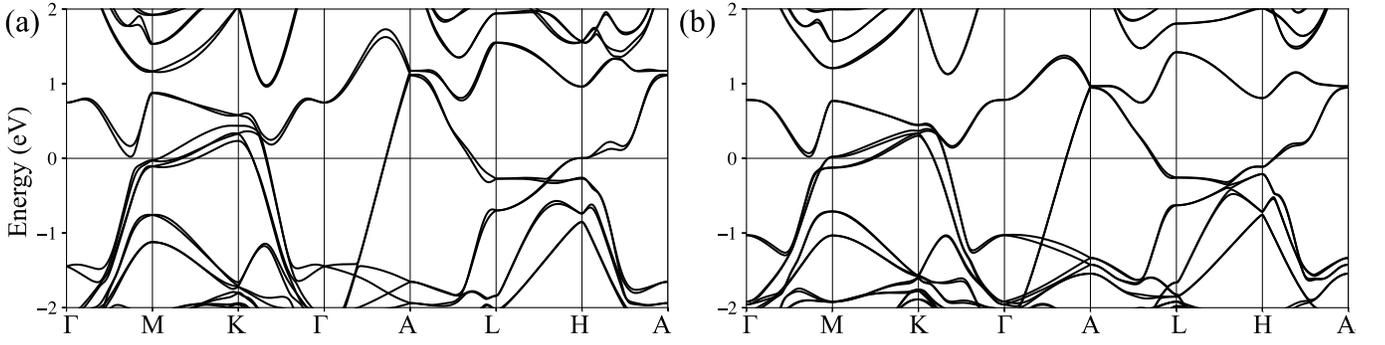}
	\caption{(a) Band structure of TaSi$_2$ along the high-symmetry lines sketched in Fig. 1(c). (b) The same as (a) for NbSi$_2$.}
	\label{fig:tanbstruct}
\end{figure*}

\begin{figure*}
	\centering
	\includegraphics[width=\textwidth]{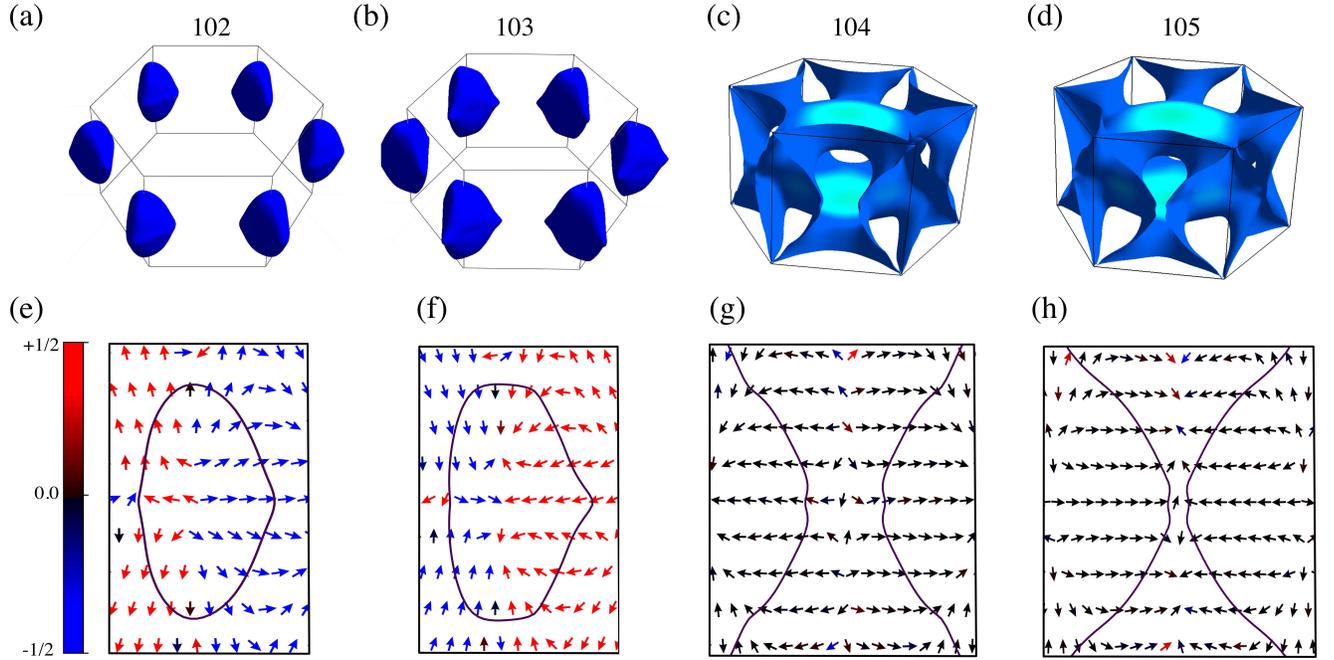}
	\caption{(a)-(d) Fermi surfaces of TaSi$_2$ calculated at the Fermi level. Bands numbered as 102 and 103 are elliopsoids centered at the $K$-points, while bands 104 and 105 are open surfaces whose projections at each of the side (rectangular) BZ faces are hyperbolic-shaped contours around the $M$-points (see panels (e) and (f)). (e) Spin texture of band 102 projected onto the side face of the BZ and centered at the $K$-point. The contour corresponds to $E=E_F$. (f) The same, plotted for band 103. (g) The same as (e)-(f) calculated for band 104 and centered around the $M$-point. (h) The same as (g) plotted for band 105. The spin textures around the $K$-points follow a purely Weyl-type SOC and those around the $M$-points reveal a Weyl-Dresselhaus SOF. The arrow colors indicate spin components normal to the planes. Fermi surfaces were visualized using FermiSurfer.\cite{fermisurfer}
  }
	\label{fig:tasi2fsspintxt}
\end{figure*}

\begin{figure*}
	\centering
	\includegraphics[width=\textwidth]{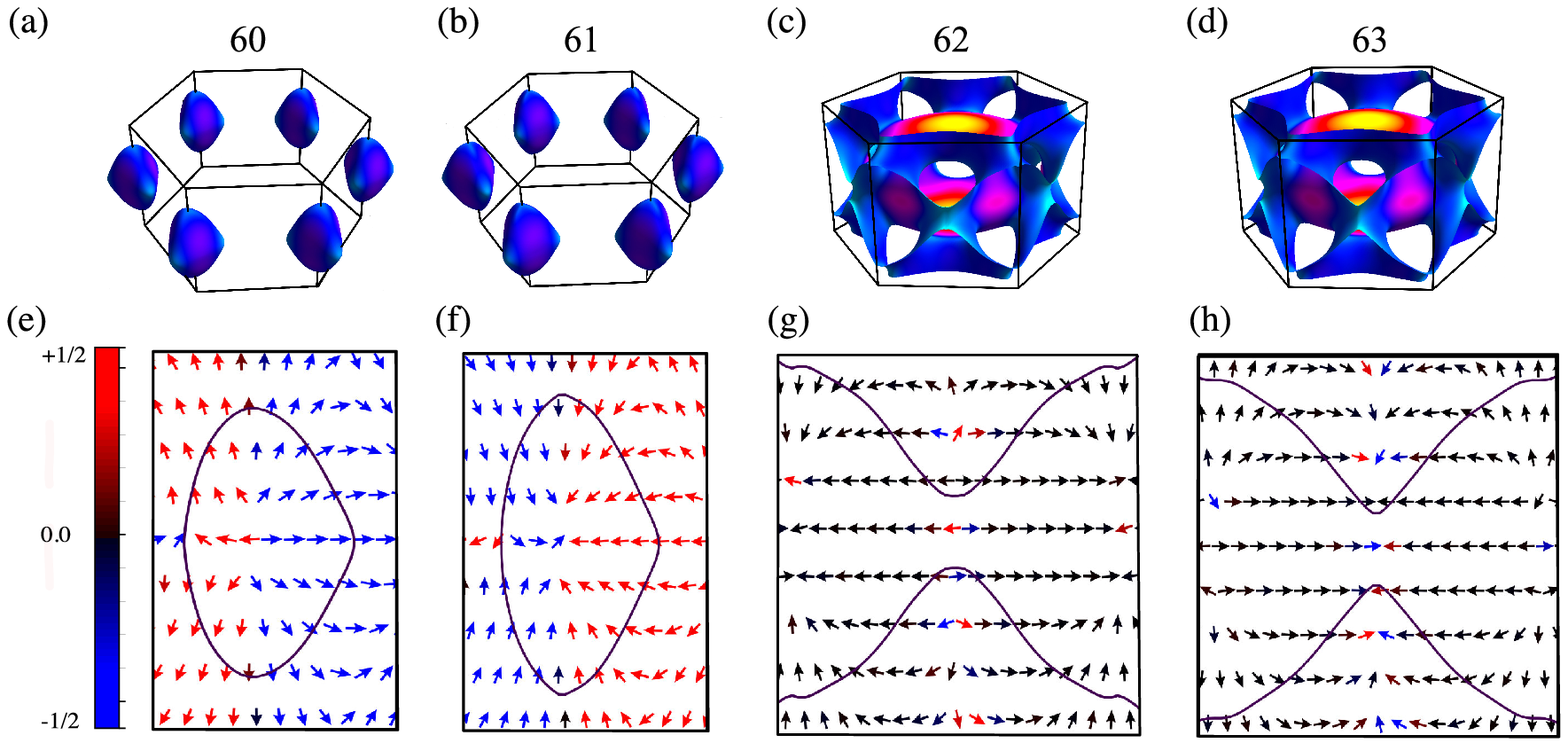}
	\caption{The same as Fig. \ref{fig:tasi2fsspintxt}, but calculated for NbSi$_2$.}
	\label{fig:nbsi2fsspintxt}
\end{figure*}

\textit{Structure and electronic properties.}
We will now consider two structurally identical metal disilicides, NbSi$_2$ and TaSi$_2$ whose right- and left-handed crystals are described by space groups P6$_2$22(180) and P6$_4$22(181), respectively. The structures, shown in Fig. \ref{fig:tanbstruct_0}, possess 6 two-fold (C$_2$) rotational symmetries with axes within the \textit{a-b} plane. Two enantiomers of the crystals can be distinguished by the opposite helicities of the pairs of X-Si-M (X = Ta/Nb) chains that run along the \textit{c-}axis, as illustrated in Fig. \ref{fig:tanbstruct_0} (b). The band structures calculated for TaSi$_2$ and NbSi$_2$ are displayed in Fig. \ref{fig:tanbstruct}(a)-(b), respectively, and they confirm that both materials are Weyl semimetals, in line with the previous studies.\cite{anisotropic, chiral_structure_jpsj, arunesh_npj, zhang2018topological} The Fermi surfaces, shown in Fig. \ref{fig:tasi2fsspintxt}(a)-(d) and Fig. \ref{fig:nbsi2fsspintxt}(a)-(d), respectively, consists of two pairs of spin-splitted Fermi sheets. We note that the splitting energies in TaSi$_2$ are over twice larger than in NbSi$_2$, in agreement with the previous calculations and the measurements of de Haas-van Alphen effect.\cite{chiral_structure_jpsj} Such a result is not surprising, as the primary contributors to SOC in these materials are Ta-5\textit{d} and Nb-4\textit{d} electrons. We can also observe that the band splitting at $E_F$ is the largest close to the high symmetry points $M$ and $K$, indicating that the influence of SOC on the bands varies significantly across the BZ.

\begin{figure*}
	\centering
	\includegraphics[scale=1]{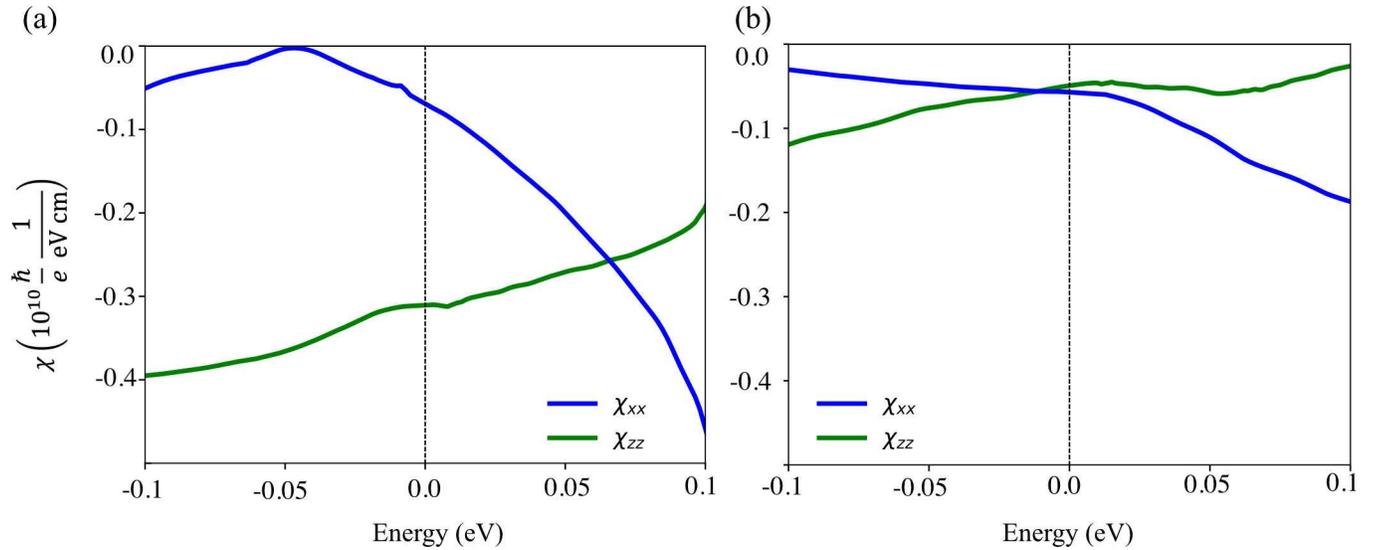}
	\caption{(a) CISP tensor of left-handed TaSi$_2$ calculated as a function of the chemical potential. Different tensor components are distinguished by colors. (b) Same as (a) calculated for the left-handed NbSi$_2$. The components $\chi_{yy}$ are omitted, as they are equal to $\chi_{xx}$.}
	\label{fig:tanbcisp}
\end{figure*}

\textit{Spin textures.} The spin textures corresponding to these Fermi sheets are visualized in Fig. \ref{fig:tasi2fsspintxt}(e)-(h) and Fig. \ref{fig:nbsi2fsspintxt}(e)-(h) for TaSi$_2$ and NbSi$_2$, respectively. For convenience, we projected them onto the planes (side faces of BZ) centered at $K$ and $M$, similarly as in the case of Te and Se. To understand the spin patterns, we performed the group theory analysis of the high symmetry points using the BCS server. We found that $\Gamma$, $A$, $H$, $K$ are described by either D$_6$ or D$_3$ point group symmetry which only allows for Weyl-type spin texture.\cite{zunger}  Two other points, $M$ and $L$, are described by little point group D$_2$, thus they can host a combination of Weyl and Dresselhaus SOC.\cite{zunger} These findings agree with the first-principles results; the spin patterns around the $K$-point are radial, alternating between outward and inward orientation for two subsequent bands 102 and 103. Also the spin texture of the band centered at the $M$-point is predominantly radial (Fig. \ref{fig:tasi2fsspintxt}(e)-(f) and Fig. \ref{fig:nbsi2fsspintxt}(e)-(f)), with tangential components visible only close to the corners of the plot for TaSi$_2$ (Fig. \ref{fig:tasi2fsspintxt}(g)-(h)) and around the $M$-point for NbSi$_2$ (Fig. \ref{fig:nbsi2fsspintxt}(g)-(h)).

\textit{Current-induced spin accumulation.} The experimental studies of chirality-dependent spin transport in metal disilicides revealed spin signals due to the electric current flowing along the chiral axis.\cite{japanese_prl} In our recent study, we suggested that such results could be attributed to the collinear REE occurring in TaSi$_2$, due to the presence of dominant Weyl-type SOF.\cite{arunesh_npj} However, similarly to the elemental semiconductors, the REE tensor has three (two independent) components $\chi_{xx} = \chi_{yy}$ and $\chi_{zz}$,\cite{karma_arxiv} and the former were not investigated experimentally.

The magnitudes of different components of $\chi$ tensor calculated as a function of the chemical potential are plotted in Fig. \ref{fig:tanbcisp}(a)-(b) for TaSi$_2$ and NbSi$_2$, respectively. Surprisingly, in TaSi$_2$ we observe an even higher value of $\chi_{xx}$ than $\chi_{zz}$ at the Fermi level, and in NbSi$_2$ these two components are nearly equal. This result can be rationalized based on the spin textures shown in Fig. \ref{fig:tasi2fsspintxt} and Fig. \ref{fig:nbsi2fsspintxt}; we can notice that the radial spin patterns are rather isotropic. They differ from the spin texture of Te, where the spins were almost entirely oriented along the screw axis, leading to the enhancement of the $\chi_{zz}$ component. Moreover, the band velocities are high in all directions, which may explain why the values of REE at $E_F$ are nearly two orders of magnitude lower than in Te.

Both disilicides have comparable REE, because their Fermi surfaces and the corresponding spin textures which give rise to the spin accumulation are overall quite similar. Differences in the magnitudes of $\chi$ can be attributed to the fact that the spin-splitting of the bands is larger in TaSi$_2$ than in NbSi$_2$. This can be noticed in Fig. 7(a)-(b), close to $E_F$ and along the crystallographic directions $\Gamma M$, $AH$, $AL$ and $MK$ that determine the magnitudes of $\chi_{xx}$ and $\chi_{yy}$ tensor components. The weaker SOC in NbSi$_2$ gives rise to nearly degenerate bands, and their contributions to the effective spin accumulation nullify each other, similarly as in the case of Se. Our results for chiral disilicides are thus consistent with the conclusions drawn for the elemental semiconductors.

\section{Conclusions}
In conclusion, we studied the chirality-dependent collinear REE in a few representative materials, focusing on the role of SOC. We found that the unconventional spin accumulation is favorable in crystals that host purely Weyl (radial) SOF giving rise to parallel spin-momentum locking. We observed that the magnitude of the induced spin density is sensitive to both the spin polarization of bands in the momentum space and band velocities. For example, the bands of Te are highly spin-polarized along the screw axis, which implies the highest spin accumulation along this direction, whereas the spin texture is more isotropic in Se, leading to almost equal magnitudes of $\chi$ along different axes. The values, as a function of the chemical potential, are additionally modulated by the band velocities; low velocity may increase the spin accumulation generated along a specific axis, even when the band is weakly spin-polarized in this direction.

The effect of the strength of SOC is more subtle to analyze.\cite{taiwan} On one hand, the spin-orbit interaction lifts both energy and spin degeneracy of non-centrosymmetric crystal structures and it is a necessary condition for the emergence of REE.\cite{karma_arxiv} On the other hand, the high SOC does not per se lead to large magnitudes of REE; note that two chiral disilicides with different strengths of SOC show similar $\chi$ around $E_F$. Our results reveal that stronger SOI provides larger spin-splitting of the bands and often prevents compensating the spin accumulation that originates from oppositely spin-polarized bands. It can therefore indirectly increase the REE, but only in the energy regions where the band splitting is significant.

Our research on the collinear REE in chiral materials has practical applications in the design of spintronic devices that use spin-to-charge conversion in unconventional configurations. Additionally, it provides a valuable reference for computational studies searching for efficient CSC materials. Even though it is difficult to predict the magnitude of REE without calculating the electronic structure, it will guide high-throughput simulations and facilitate their analysis. Finally, we clarify the role of SOC which is believed to be crucial for observable REE in solid-state materials, and we highlight the significance of flat bands that may contribute to the effect. This can be relevant for future studies of CISS which occurs also in systems with weak SOC.

\begin{flushright}

\end{flushright}


%

\end{document}